# SD-AREE: A New Modified Caesar Cipher Cryptographic Method Along with Bit-Manipulation to Exclude Repetition from a Message to be Encrypted

Somdip Dey (Author)
B.Sc. (Hons)
Department of Computer Science
St. Xavier's College [Autonomous]
Kolkata, West Bengal, India.

*Abstract* – **In this paper the author presents a new cryptographic technique to exclude the repetitive terms in a message, when it is to be encrypted, so that it becomes almost impossible for a person to retrieve or predict the original message from the encrypted message. In modern world, cryptography hackers try to break a code or cryptographic algorithm or try to retrieve the key, which is needed to encrypt a message, by analyzing the insertion or presence of repetitive bits / characters (bytes) in the message and encrypted message to find out the encryption algorithm or the key used for it. So it is must for a good encryption method to exclude the repetitive terms such that no trace of repetitions can be tracked down. For this reason we apply SD-AREE cryptographic method to exclude repetitive terms from a message, which is to be encrypted. In SD-AREE method the repetitive bits / characters are removed and there is no trace of any repetition in the message.**

Keywords – Caesar Cipher; Cryptography; Repetition exclusion; Encryption; Decryption;

I. INTRODUCTION

In today's world, security is a big issue and securing important data is very essential, so that the data can not be intercepted or misused for illegal purposes. For example we can assume the situation where a bank manager is instructing his subordinates to credit an account, but in the mean while a hacker interpret the message and he uses the information to debit the account instead of crediting it. Or we can assume the situation where a military commander is instructing his fellow comrades about an attack and the strategies used for the attack, but while the instructions are sent to the destination, the instructions get intercepted by enemy soldiers and they use the information for a counter-attack. This can be highly fatal and can cause too much destruction. So, different cryptographic methods are used by different organizations and government institutions to protect their data online. But, cryptography hackers are always trying to break the cryptographic methods or retrieve keys by different means and one of such methods include the process of inclusion of repetitive texts or characters in a message and then encrypt them to study the cryptanalysis of the method and retrieve the key, which is needed for cryptographic method, or break the algorithm. SD-AREE is a cryptographic method to exclude the repetitive characters in a message to be encrypted and this technique is a type of symmetric key cryptography.

The modern day cryptographic methods are of two types: (i) symmetric key cryptography, where the same key is used for encryption and for decryption purpose. (ii) Public key cryptography, where we use one key for encryption and one key for decryption purpose.

Symmetric key algorithms are well accepted in the modern communication network. The main advantage of symmetric key cryptography is that the key management is very simple. Only one key is used for both encryption as well as for decryption purpose. There are many methods of implementing symmetric key. In case of symmetric key method, the key should never be revealed / disclosed to the outside world or to other user and should be kept secure.

SD-AREE is also a symmetric key cryptographic method and the best part of this method is that this method can be used with other cryptographic method to make those encryption strong enough to be broken by hackers.

In this paper the authors present the cryptographic technique SD-AREE, which is a both bit level and byte level cryptographic method. First the message is broken up into bits and it is saved in a matrix then Matrix –cycling Operation is performed on that matrix for random number of times. Then we apply a modified form of Advanced Caesar Cipher Cryptographic Method [11]. In cryptography, a Caesar cipher, also known as a Caesar's cipher or the shift cipher or Caesar's code or Caesar shift, is one of the simplest and basic known encryption techniques. It is a type of replace cipher in which each letter in the plaintext is replaced by a letter with a fixed position separated by a numerical value used as a "key". Caesar Cipher is or was probably the very first encryption methodology. It is a type of substitution cipher in which each letter in the plaintext is replaced by a letter some fixed number of positions down the alphabet. For example, with a shift of 3,A would be replaced by D, B would become E, and so on.

In SD-AREE, we first perform cyclic operation on the matrix, where each character (for ASCII encoded text) or byte is saved in bit wise and then extract the ASCII value of each byte (character) of the text after cyclic operation, and then we add two numbers with it, which we generate randomly from the pass-key (symmetric key) provided for encryption, to generate a new ASCII value and print out the character of that ASCII value. The two numbers, which are generated from the pass-key, are generated totally randomized and it is impossible to predict the pass-key for that reason. Again, one of the two numbers is generated using a random polynomial function and it changes every time with the byte, of which ASCII value it is being added to. For this reason, it is almost impossible to predict the pattern of the cipher method or the key in case the hackers want to retrieve the data.

In SD-AREE we take the symmetric key as password given by the user and from that key we generate unique codes, which are successively used to encrypt the message. The methods used in SD-AREE to encrypt the message can be reverse engineered to decrypt the same message by providing the key (password).

## II. ENCRYPTION

### a) Generation of Code and power_ex from the Symmetric Key

The key is provided by the user in a string format and let the string be 'pwd[]'. From the
given key we generate two numbers:
'code' and 'power_ex', which will be used for encrypting the message. First we generate the 'code' from the pass key. Generation of code is as follows:

To generate the code, the ASCII value of each character of the key is multiplied with the string-length of the key and with $2^i$, where 'i' is the position of the character in the key, starting from position '0' as the starting position. Then we sum up the resultant values of each character, which we got from multiplying, and then each digit of the resultant sum are added to form the 'pseudo_code'. Then we generate the code from the pseudo_code by doing modular operation of pseudo_code by 16, i.e.
    code = (pseudo_code Modulus 16).
    If code==0, then we set code =pseudo_code

The Algorithm for this is as follows:
Let us assume, pwd[] = key inserted by user
$pp = 2^i$, $i=0,1,2,\ldots\ldots n; n \in N$.
Note: i can be treated as the position of each character of the key.
Step 1: p[] = pwd[]
Step 2: pp = $2^i$
Step 3: for (i=0;i < strlen(pwd); i++)
    p[i] = pwd[i];
    p[i] = p[i] * strlen(pwd) * pp;
    csum = csum + p[i];
Step 4: while (csum ≠ 0)
    c = csum Modulus 10;
    pseudo_code=pseudo_code +c;
    csum = csum / 10;
Step 5: code = (pseudo_code Modulus 16);
Note: strlen(pwd) is string-length of the pwd[] (key).

Generation of power_ex is as follows:
Now, we generate power_ex from the pseudo_code generated from the above step. We add all the digits of the pseudo_code and assign it as temporary_power_ex. Then we do modular operation on temporary_power_ex with 3 and save the resultant as power_ex.
i.e.
power_ex = (temporary_power_ex Modulus code)
If power_ex == 0 OR power_ex == 1, then we set power_ex = code.

For example, if we choose the password, i.e. the key to be 'hello world'. Then,
Length of pwd = 11
code = 10
power_ex = 4

Thus, we generate code and power_ex from the key provided by the user.

### b) Bit Level Matrix-Cyclic Operation

Now, after the generation of 'code' and 'power_ex', the text, which needs to be encrypted, is taken up as a string and each charcter (for ASCII encoded text) or byte is broken up into bits (8-bit pattern). Then a matrix table is formed, which is of dimension (8 x 8), provided that the number of bytes is equal or more than 8, and keep on forming such matrices until all bytes of the message to be encrypted have been transformed into bit-wise matrix format. If the number of bytes is less than 8 then the matrix formed will have the dimension (m x 8), where 'm <=8'. In this matrix, we save the bits of each byte in the column of the matrix and save each byte in the row of the matrix. For example: if the text to be encrypted is 'aBc', then it can be represented in matrix form after extraction of bits in the following fashion:

| Matrix (3 x 8) | Bit 7 (MSB) | Bit 6 | Bit 5 | Bit 4 | Bit 3 | Bit 2 | Bit 1 | Bit 0 (LSB) |
|---|---|---|---|---|---|---|---|---|
| a -> | 0 | 1 | 1 | 0 | 0 | 0 | 0 | 1 |
| B -> | 0 | 1 | 0 | 0 | 0 | 0 | 1 | 1 |
| c -> | 0 | 1 | 1 | 0 | 0 | 0 | 1 | 1 |

Note: In the above example matrix we can see that the number of bytes is less than 8, so only 1 matrix is formed and the dimension of the matrix is (3 x 8), where m=3, which is <8.

Now, after the matrices are created we perform cyclic operation on the matrix several times, i.e. multiple times and form a new set of values.

Matrix-cyclic operation can be explained in the following example:

| A | B | C | D |
|---|---|---|---|
| L | M | N | E |
| K | P | O | F |
| J | I | H | G |

Fig 1.1: Real Matrix

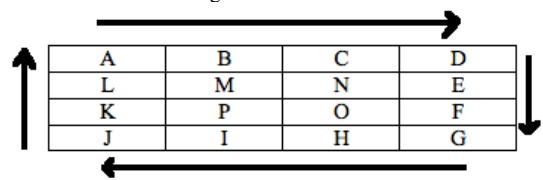

Fig 1.2: Cyclic Operation Starts

| L | A | B | C |
|---|---|---|---|
| K | N | O | D |
| J | M | P | E |
| I | H | G | F |

Fig 1.3: Matrix After Cyclic Operation

We perform the cyclic operation 'n' number of times, where n= code, which is generated from the key. After the cyclic operation we again change back the binary value of the bytes into its decimal form i.e. the ASCII form.

Algorithm for this whole step:
Step 1: extract all bytes of the message into a string
Step 2: extract the bits of each byte from the string i.e. convert ASCII value of each byte to binary form
Step 3: save the bits in a 2 dimension array such as a[r][c], where r -> row and c -> column
Step 4: r = each byte
    c = each bit of that byte
Step 5: perform cyclic() 'n' times on a[r][c], where n = code

Step 6: convert the binary form to ASCII value (decimal form)
Step 7: store the result in text[]

Note: text[] is an array, where the ASCII value of the bytes after bit level encryption is stored.

### c) Encrypting the Bit-Message using code and power_ex (Modified Caesar Cipher Method)

Now we use the code and power_ex, generated from the key, to encrypt the bit level encrypted text. We extract the ASCII value of each character of the text, which is produced after bit level encryption, and add the code with the ASCII value of each character. Then with the resultant value of each character we add the $(power\_ex)^i$, where i is the position of each character in the string, starting from '0' as the starting position and goes up to n, where j=position of end character of the message to be encrypted, and if position = 0, then $(power\_ex)^i = 0$. It can be given by the formula:

$$text[i] = text[i] + code + (power\_ex)^i$$

If, ASCII value of text[i] > 255, then set
text[i] = (text[i] Modulus 255)
Note: 'i' is the position of each character in the text and text[] is the message to be encrypted, where text[i] denotes each character of the text[] at position 'i'.

For example, if the text to be encrypted is 'aaaa' and key=hello world, i.e. text[]=aaaa and pwd = hello world, then,
$a^0$ -> 97 + 10 + 0 = 107 ->k
$a^1$ -> 97 + 10 + 4 = 111 ->o
$a^2$ -> 97 + 10 + 16 = 123 ->{
$a^3$ -> 97 + 10 + 64 = 171 -> «
where 0-3 are the positions of 'a' in text[]
(as per formula given above)
The text 'aaaa' becomes 'ko{«' after execution of this Step.

Since, the value of $(power\_ex)^i$ increases with the increasing number of character (byte) i.e. with the increasing number of string length, so we have applied the method of **Modular Reduction** [9][10] to reduce the large integral value to a smaller integral value.

To apply Modular Reduction we apply the following algorithm:
Step 1: n = power_ex * code * 10 //generate a random number 'n' from code and power_ex
Step 2: calculate $n^{th}$ prime number
Step 3: For (i=0; i<string length[text]; i++)
Step 4: $(power\_ex)^i$ Modulus ($n^{th}$ prime number)
Step 5: replace the value of $(power\_ex)^i$ with the value generated in Step 4
Step 6: Next i

Following the above step, we can reduce the value of $(power\_ex)^i$ to a significantly smaller usable number.

### III. DECRYPTION

The above processes of encryption of SD-AREE method can be reverse engineered to get back the original text (message) and thus decryption of the encrypted text can be executed. In case of decryption of matrix operation of cycling method, the cycling process can be reversed, i.e. the cycling method is executed in reverse direction, to get back the original text.

The main formulae used for decryption is:

$$text[i] = text[i] - code - (power\_ex)^i$$

Note: If, ASCII value of text[i] < 0, then set
text[i] = (text[i] Modulus 255)
'i' is the position of each character in the text and text[] is the message to be encrypted, where text[i] denotes each character of the text[] at position 'i'.

### IV. RESULTS AND DISCUSSIONS
In the following table few results are given:

| Text to be Encrypted | Encrypted Text |
|---|---|
| aaaaa | @X¢Ù$ |
| 111aaaaaaaa111 | Ù@rÉÛ`ù~o®?„Ô* |
| aaaaaaaaaaaaa | Óš²©Ú <br> pÙM> oÄ i |
| aabbccdd | @X²ÈÚâp[¥ |
| he is good | D_²šãä1ù5ÐÔ |
| 11fff11 | (ztÎÆùb |
| ÿÿÿÿÿÿÿÿÿ | \|*o,s}~ox |
| St. Xavier's is an autonomous college | {  szzon*~y*~ <br>       kuo*  z* <br> pÙM> |
| January 16, 2012 is the 152 Foundation Day of St. Xavier's. Fr. Principal joined the staff and students in a small but meaningful celebration. | 5Y¶Ìñ  {Ù5{  4{ú'  aM / ðÏ ⋯ 2Îfá8ü† t_‰œ¼ R  ˆ÷iÀò  ¶ xpw0‹C\  :q\'Óš Jx  ~u~ˆ  0s‡ò˜ªFh, 0x}`Lp*aLPÊ:hqnenxA4 ]p c w |

From the above results, it can be seen that, all the repetitive terms are excluded from the encrypted text and it can never be figured out just from the encrypted text that there was any repetition in the text message. SD-AREE is a provable good method to exclude repetitive terms, but, it is also evident from the results shown above, that SD-AREE is not a very good cryptographic method to encrypt a full text, i.e. it is not a full proof encryption technique. And this problem can be avoided by adding other encryption techniques with this method. For this reason this method is meant to be used with other cryptographic methods as a cryptographic module in order to make those methods strong.

**Discussions and Limitations of SD-AREE:**
SD-AREE method should never be treated as a lone method for encryption, but this method is a cryptographic method, which exclude the repetitive characters from the text message, which is to be encrypted. To make SD-AREE a fully strong encryption technique, other encryption methods should be added with SD-AREE method.

In most of the recent encryption techniques, symmetric key, which is used for encryption, is mostly 16 –bits long. So, for SD-AREE method, if a 16-bit key is provided for encryption and if the ASCII value of the bit is 255, then,

$$csum = \sum_{i=1}^{16} 255 * i * (2)^i$$
$$= 2404388880,$$

then, code = 45 and
power_ex = 9 [According to SD-AREE method]

And if the number of bits in the symmetric key increases, then there is a possibility that the value of power_ex may increase too, and this is a problem. If the value of power_ex is very big then the value of (power_ex)$^i$ will be even larger, and since this value is used in SD-AREE encryption method, so the value may be very large to store in memory. This problem may arise if SD-AREE method is used to encrypt large text files or any other type of large files.

So, to do away with this problem, we have included the step in 'power_ex' generation:
power_ex = (temporary_power_ex Modulus code);
If power_ex == 0 OR power_ex == 1, then we set power_ex = code.

Note: To do away with the above stated problem the author have provide this provision; although, this problem can be handled by using modular reduction to reduce the large integral value to suitable form to compute the SD-AREE method.

But, this provision may be changed as per user or the cryptographer, and is always welcomed to change this part as needed.

## V. SPECTRAL ANALYSIS OF FREQUENCY OF CHARACTERS USING SD-AREE

One of the classical cryptanalysis method used is by detecting the frequency of characters in the encrypted text (message). So to test the effectiveness of SD-AREE method, spectral analysis of the frequency of characters are closely observed.

Using SD-AREE method we ran many analysis and tested different strings as input and used various methods of cryptanalysis. To show the usefulness and integrity of this cryptographic module, we used spectral analysis of the frequency of characters.

| Message | Encrypted Message |
|---|---|
| aaaaaaaabbbbaaaaaaaa | @X²ÉÛâp[Ÿ? "   "([`E    ð |

Table 1.1: Encrypting Palindrome using SD-AREE

In Fig 2.2 we show the spectral analysis of the string of 'aaaaaaaabbbbaaaaaaaa'.

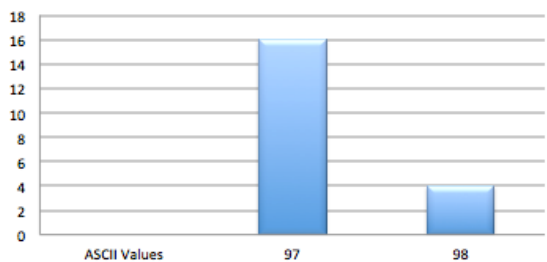

Fig 2.1: Spectral Analysis of frequency of characters of string 'aaaaaaaabbbbaaaaaaaa'

After using SD-AREE method we encrypted the string ' aaaaaaaabbbbaaaaaaaa ', which is a palindrome, and as an output we got ' @X²ÉÛâp[Ÿ? " "([`E    ð ', which is the encrypted string after using SD-AREE method (shown in Table 1.1). Fig 2.2 shows the spectral analysis of the frequency of characters of the encrypted string. From the figures it is evident that SD-AREE method is very effective to exclude repetition and provide no trace of that in the encrypted text.

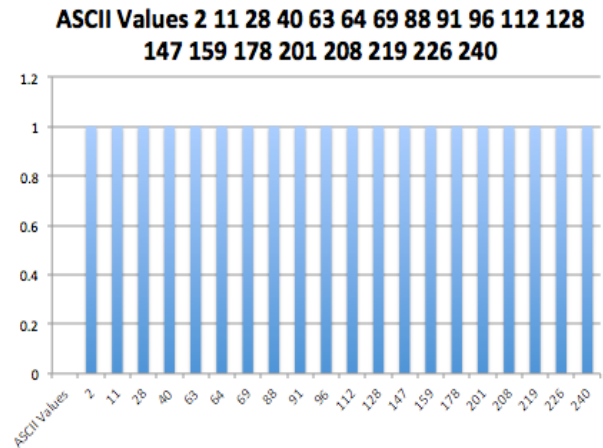

Fig2.1: Spectral Analysis of Frequency of Characters of the Encrypted String of the Palindrome

| Message | Encrypted Message |
|---|---|
| aaaabbbbccccddddeeeefffff | @X²ÉÝão^®?" ´<3\À  ÙŸµ_^‰ê |

Table 1.2: Encrypting a Ascending String using SD-AREE

As another test case, ' aaaabbbbccccddddeeeefffff ' is taken as input string and the encrypted string using SD-AREE method is '@X²ÉÝão^®?"   ´<3\À  ÙŸµ_^‰ê '. Encryption using SD-AREE shown in Table 1.2. Fig 2.3 shows the spectral analysis of frequency of characters of the string ' aaaabbbbccccddddeeeefffff ' and Fig 2.4 shows the spectral analysis of frequency of characters of the encrypted string.

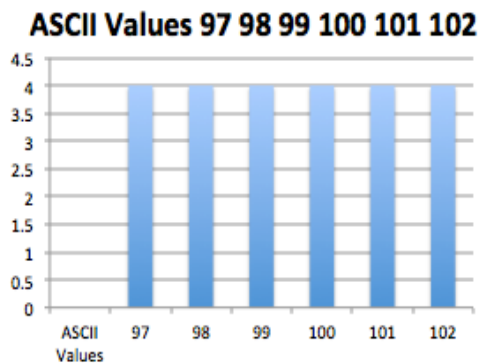

Fig 2.3: Spectral Analysis of Frequency of Characters of the String ' aaaabbbbccccddddeeeefffff '

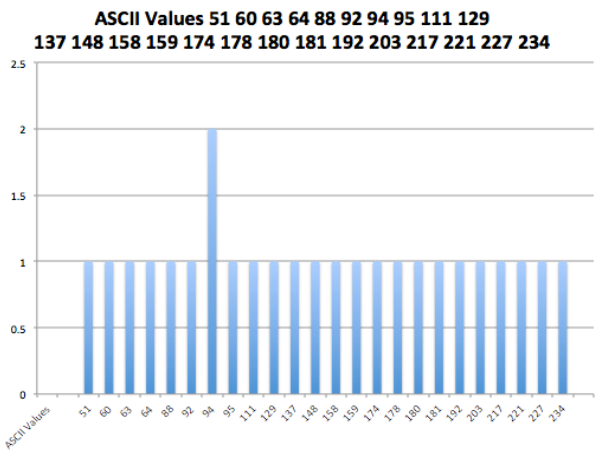

Fig 2.4: Spectral Analysis of Frequency of Character of Encrypted String

Thus from the spectral analysis of the SD-AREE method, it is evident that SD-AREE is successful to exclude the evidence of repetitive terms (bytes) from a message, which is needed to be encrypted.

## VI. GENERAL CRYPTANALYSIS

If we analyze the above SD-AREE method, then we can see that the use of polynomial function in the method have significantly increased the strength of encryption. Not only this, but the inclusion of bit-level encryption along with byte-level encryption has also increased the security of the cryptographic method. Although this technique is based on the Caesar Cipher method but modifying Caesar Cipher by introducing random polynomial function and modular reduction have made a strong cryptographic method. This helps the encrypted text to be almost impossible to be detected by including repetitive bytes or characters and it also makes the method strong against Differential Attack (Differential Cryptanalysis).

## VII. CONCLUSION

SD-AREE method is meant to be used with other encryption techniques present in the world, so that those encryption techniques can be made stronger and the possibility of breaking those encryption techniques or retrieving the key can be avoided by excluding the evidence of repetitive terms from the message, which is to be encrypted. In this way we can avoid the possibility of cryptanalysis and differential cryptanalysis (differential attack) occurring for an encryption technique.

## ACKNOWLEDGMENT

Somdip Dey expresses his gratitude to all his fellow students and faculty members of the Computer Science Department of St. Xavier's College [Autonomous], Kolkata, India, for their support and enthusiasm. He also thanks Dr. Asoke Nath, professor and founder of Computer Science Department of St. Xavier's College (Autonomous), Kolkata, for his constant support and helping out with the preparation of this paper.